\documentstyle[prl,multicol,aps]{revtex} 
\begin{document}
\draft
\twocolumn
%\title
\begin{center}
{Random Energy Model as a paradigm of complex systems}\\
%\author
{D.B. Saakian}\\
%\address
{Yerevan Physics Institute,
Alikhanian Brothers St. 2, Yerevan 375036, Armenia.}
\end{center}
%\maketitle

\begin{abstract}
A quadratic extension of REM has been treated. Discussed here is the
origin of relation of REM to strings and other complex physical phenomena.
Two basic features of the REM class of complex phenomena were identified:
the double thermodynamic reflection (a hierarchy of free energies) including
the strong reflection at the upper level (the free energy on the order of a
logarithm of the degrees of freedom) and the loss (complete or partial) of the
local symmetry property. Two main classes of complex phenomena related to REM
are seen: the spin glass phase of REM and the boundary the spin glass-ferromagnetic
phases. Some examples of physics interest are analyzed from this viewpoint.
\end{abstract}

\vspace{5mm}
While quite a number of physical theories (with the elements of probability
theory) have been successfully treated lately, there still remain some unsolved
 problems, the example of which are strings, turbulence, spin glasses, 2d
 quantum disorder. The last three ones refer to the (loosely defined ) fields of physics of complex systems. One can also mention important biophysical systems with complex behavior such as the protein folding, the evolution. It is noteworthy that however
 different these 6 fields seem, they are associated with REM[1]. It seems reasonable, therefore, to establish their relationship to REM with a view to better understanding of these theories, as well as to comprehend the complex phenomena and, if possible,
 to specify their boundaries, using REM as a paradigm of complex systems.\\
What is the main feature of REM that makes for its power? The human mind is envisaged to be a complex system and so the information processing must be an essential property of a complex phenomena.
From this viewpoint REM is unique-it functions as optimal coding device by saturating the Shannon limit for highest possible rate of error-free coding [2-4]. That is why the model may partially resemble even complex adaptive systems of the type of biologi
cal evolution[5].\\
Recently some interesting data on the relation of REM to strings and 2d quantum disorder [6-8] and decaying Burgers turbulence [9] have been obtained. We are intent to clarify further the relation of REM to the quantum Liouville model on the basis of REM.
\\
Now consider the standard definition of REM. We have M random energy levels distributed according to formula 
\begin{equation}
\label{e1}
P(E)=\sqrt{\frac{1}{2a\pi}}\exp (-E^2/2a)
\end{equation}
and the spin glass (SG)-paramagnetic (PM) phase transition takes place at $\beta =\sqrt{2\frac{\ln M}{a}}$ 
point. At complex temperatures $\beta\equiv \beta_1+i\beta_2$ there may exist the Lee-Yang-Fisher phase (LYF) [10],
when $\beta_1\le \frac{\beta_c}{2}$.  When $\beta$ is a purely imaginary number, the transition between LYF and PM phases occurs at $\beta_2 =\frac{\beta_c}{\sqrt{2}}$.
Thermodynamically REM is equivalent to the p-spin model in the limit of large p:
\begin{equation}
\label{e2}
H=-\sum_{1\le i_1<i_2..<i_p\le N}j_{i_1..i_p}s_{i_1}..s_{i_p}
\end{equation}
Here $j$ are the  random couplings distributed according to normal law.
For averaging of  $\ln Z\equiv \ln \sum_i e^{-\beta E_i}$ in distributions $E_i$,   Derrida used the trick:
\begin{equation}
\label{e3}
\ln z=\Gamma'(1)+\int_{0}^{\infty}\ln t d [e^{-tz}]
\end{equation}
It is easy to derive this equality for real positive $z$. Then analytically continue that for imaginary z as well. In the 
last case the real part on both the sides of (3) has to be taken.
To calculate (3), one can factorize the integrand in different
energies and transform (3) to the form
\begin{equation}
\label{e4}
<\ln Z>=\Gamma'(1)+\int_{0}^{\infty}\ln t d [f(t)^{2^N}]
\end{equation}
where $M=2^N,\lambda=\beta\sqrt{N},a=N/2$
\begin{equation}
\label{e5}
f(t)=
\frac{1}{2\pi i}\int_{-i\infty}^{i\infty}e^{  \frac{y^2\lambda^2}{4}-y\ln t}\Gamma(y) dy
\end{equation}
In (5) the integration loop goes round the point 0 from the wright. In this
representation the function $f(t)$ 
can be determined for any complex value (t).
The main contribution to  (4) in thermodynamic limit $N\to
\infty$ is given
by region $\ln t\to -\infty$, so it is possible to
consider the saddle point method in (5). The saddle point is in $[-1,0]$,
or left wards. When the integration loop is moved to encircle
the saddle point, different poles of  $\Gamma (y)$ function are intersected
in the process, by which means the phases of the model are obtained. SG
phase is obtained when only $y=0$ pole is is intersected, the Lee-Yang-Fisher
like phase (LYF) (at complex temperatures) results at intersection of
$y=-2$ pole, PM- of y=-1 pole. The intersection of poles to the left of $y=-2$ 
does not give new thermodynamic phases and leads only to singularities in finite volume corrections.\\
Now consider the normal distribution for $E_i$ with a quadratic form of general type (instead of diagonalized one in REM), and define the generating function $L(t)$ as:
\begin{eqnarray}
\label{e6}
&&L(t)=\frac{1}{\sqrt{det\{A\}}}\prod_{i=1}^{2^N}\sqrt{\frac{1}{\pi}}\int_{-\infty}^{\infty}dE_i
\nonumber \\
&&\exp\{(-\sum_{ik}A_{ik}E_iE_k)
-t\sum_i e^{\beta E_i}\}
\end{eqnarray}
In analogy with (3) one can derive
\begin{eqnarray}
\label{e7}
&&\int \ln t d L(t)=-\Gamma'(1)+\frac{1}{\sqrt{det\{A\}}}\prod_{i=1}^{2^N} \sqrt{\frac{1}{\pi}}\int_{-\infty}^{\infty}d E_i
\nonumber \\
&&e^{\{(-\frac{1}{2}\sum_{i k}
A_{i k}E_iE_k)\}}
\ln \sum_i e^{E_i\beta}
\end{eqnarray}
For complex $\beta$ we can assume  $Re \sum_i  e^{\beta E_i}$ to be the definition of free energy and deduce 
\begin{eqnarray}
\label{e8}
&&Re\int \ln t d L(i t)=Re\frac{1}{\sqrt{det\{A\}}}\prod_{i=1}^{2^N} \sqrt{\frac{1}{\pi}}\int_{-\infty}^{\infty}d E_i\nonumber\\
&&e^{\{(-\frac{1}{2}\sum_{i k}A_{i k}E_iE_k)\}}
\ln Re \sum_i e^{E_i\beta}-\Gamma'(1)
\end{eqnarray}
If case of $Re \beta=0$ , $L(t)$ is determined directly by (3) and, hence, in (8) instead of $L(it)$ one can 
consider $L(t)$.\\
 Introducing the second field instead of complex temperature, one can construct Z by means of formula:
\begin{equation}
\label{e9}
Z=\sum_{i=1}^{2^N}e^{\beta_1E_i}\cos(\beta_2 D_i)
\end{equation}
where  for the fields $D_i$ there is a normal distribution with some quadratic form. \\
By means of formulas like (8) one can calculate free energies $F_i$
\begin{eqnarray}
\label{e10}
&&F_1=\int_{-\infty}^{\infty}d E_e^{(-\frac{1}{2}\sum_{i k}A_{i k}E_iE_k)} \ln \sum _{i}e^{\beta_1 E_i}\nonumber \\
&&F_2=\int_{-\infty}^{\infty}d E_ie^{(-\frac{1}{2}\sum_{i k}A_{i k}E_iE_k)}\ln |Re \sum _{i}e^{(\beta_1+i\beta_2) E_i}|\nonumber \\
&&F_3=\int_{-\infty}^{\infty}d E_i d D_ie^{-\sum_{i k}\frac{A_{i k}E_iE_k+B_{i k}E_iE_k}{2}}\nonumber\\
&& \ln |Re \sum _{i}e^{\beta_1 E_i}\cos{\beta_2 D_i}|\nonumber \\
&&F_4=\int_{-\infty}^{\infty}d D_i e^{(-\frac{1}{2}\sum_{i k}B_{i k}D_iD_k)}\ln  \sum _{i}\cos{\beta_2 D_i}
\end{eqnarray}
trough functions $L_i$
\begin{eqnarray}
\label{e11}
&&L_1=\int_{-\infty}^{\infty}d E_ie^{(-\frac{1}{2}\sum_{i k}A_{i k}E_iE_k)-t \sum _{i}e^{\beta_1 E_i}}\nonumber \\
&&L_2=\int_{-\infty}^{\infty}d E_ie^{(-\frac{1}{2}\sum_{i k}A_{i k}E_iE_k)
-i t Re \sum _{i}e^{(\beta_1+i\beta_2) E_i}}\nonumber \\
&&L_3=\int_{-\infty}^{\infty}d E_i d D_ie^{-\sum_{i k}\frac{A_{ik}E_iE_k+
B_{i k}D_iD_k}{2}-it \sum _{i}e^{\beta_1 E_i}\cos\beta_2 D_i}\nonumber\\
&&L_4=\int_{-\infty}^{\infty}d D_i e^{-\frac{1}{2}\sum_{i k}B_{i k}D_iD_k-t \sum _{i}\cos\beta_2 D_i}
\end{eqnarray}
When the matrix $A$ is large, $|A|\to \infty$, then $L_i$ and
$F_i$ should have the same singularities (with respect to relevant coupling constants). It is not a direct correspondence, but rather  resembles some variant of duality, when the free energy in one model is expressed through the partition in another by me
ans of linear operations. $F_i$ models resemble REM.\\
When as A the Laplacian is taken,then $L_1$ corresponds to the quantum Liouville  model, $L_4$ to Sine Gordon,
$L_2,L_3$ are two natural extensions of the Liouville model, describing the strings in $d>1$ dimensions.\\
An analogous case of generalized random energy model (GREM) assumes rigorous solution. Here a hierarchic tree is considered and energy configurations are at the end points of the tree. Each point is connected with the root of tree 
via single trajectory. The normally distributed random variables are defined on branches of tree. The energy of 
configuration is determined as a sum of all random variables on the trajectory from the root to the endpoint corresponding 
to given configuration. In case of GREM the thermodynamic is determined by distribution function for one and two levels of energy. So, we can try to relate to our $F_i$ the GREM with distribution 
$\rho_1(\epsilon)\equiv <\delta(\epsilon-H(s,j))>$
 $\sim\exp\{-\frac{\epsilon^2}{2A^{-1}_{ii}}\}$ 
and  $\rho_2(\epsilon)\equiv <\delta(\epsilon-H(s^1,j)+H(s^2,j))>$
$\sim\exp\{-\frac{\epsilon^2}{2[A^{-1}_{ii}+A^{-1}_{jj}-2A^{-1}_{ij}]}\} $ for two configurations. It turns out at the consideration of $\rho_2$, that the physics on the whole is determined by the character of function 
\begin{equation}
\label{e12}
s(v)=\ln \sum_j \delta(\frac{ A^{-1}_{ij}}{ A^{-1}_{ij}}-v)
\end{equation}
Consider the case, when the diagonal elements $ A^{-1}_{ii}$ are independent of i. According to Derrida for monotone function $s'(v)$ 3 cases are possible. When $s'(v)$ is constant, then the 
 model is thermodynamically equivalent to REM.\\
Now let us consider the Liouville model in 2d [6]. For $L(t)$ we have 
\begin{equation}
\label{e13}
\int D(\phi(x)exp\{-\frac{1}{2}\int d^2x \phi(x)'_\alpha\phi(x)'_\alpha
-t\int d^2xe^{\beta \phi(x)}\}
\end{equation}
 The coordinate $\vec x$ serves an an  index, $\phi(x)$ are analogous to energy levels $E_i$ and the Laplacian is used for our quadratic form. Let us assume the  infrared and ultraviolet cutoffs as $a ,L$.
The number of configurations (of different $\phi(x)$ is $\frac{L^2}{a^2}$. At the calculation for the diagonal element of Green function 
\begin{equation}
\label{e14}
<\phi(\vec x)\phi(\vec x')>\equiv G(x,x')=\frac{1}{2\pi}\ln \frac {L}{|\vec x-\vec x'|}
\end{equation}
we put $|\vec x-\vec x'|=a$. The hierarchy level v stands $\frac{\ln |\vec x|-\ln a}{\ln L}$.
It is easy to find, that $s(v)=2 v$. It is, thus, in line with the case of REM according to classification of GREM by Derrida [6,7]. In such a case, in the thermodynamic limit  one can consider a simple REM with $a=G(0,a)$ and $M=\frac{L^2}{a^2}$ (see for
mula(1)) and directly obtain for the critical point
\begin{equation}
\label{e15}
\beta_c=\sqrt{\frac{4\pi \ln M}{\ln L-\ln a}}=2\sqrt{2\pi}
\end{equation}
The correct behavior  of $s(v)$ is necessary, but not sufficient condition  for coincidence of the thermodynamics of our model (13) in real space with GREM in case of $s'(v)=const$  [11]. The authors [6] had highly sound reasons (in favor of above point)-
the coincidence of Green functions in both the theories. Though subsequently they made an effort to strengthen the arguments, in reality the new arguments were no more sound, than that given in [6].\\
Now by analogy with (5) we shall try to give another representation for $L(t)$
\begin{equation}
\label{e16}
L(t)=\int_{-i\infty}^{i\infty}D x e^{\{\frac{1}{2}\sum_{i k}\hat A_{i k}^{-1}x_ix_k)-\ln t  \sum _{i} x_i\}}\prod_i\frac{\Gamma(x_i)}{2\pi}
\end{equation}
Unfortunately I failed to simplify this expression. \\
Let us consider $L_4$ (with A chosen as Laplacian). It corresponds to the to Sine-Gordon model [13]. As it is dual to $F_4$, we directly find for the transition point 
\begin{equation}
\label{e17}
\beta_c=\frac{\sqrt{\frac{2 \ln M}{\ln L-\ln a}}}{\sqrt{2}}=2\sqrt{\pi}
\end{equation}
It is connected with PM-LYF transition at $\beta_2\equiv \frac{\beta_c}{\sqrt{2}}$.
Now consider the Liouville model at complex temperatures. If we analytically continue  the formula of DDK
 \begin{equation}
\label{e18}
\beta=\beta_c\frac{\sqrt{25-d}-\sqrt{1-d}}{\sqrt{24}}
\end{equation}
into the range $d\ge 1$, then at $d=1$ we shall have PM-SG transition.\\
 For $d>1$ $\beta$ becomes a complex quantity. A possible candidate for the analytically continued Liouville model is $L_2$.  In the dual model  $F_2$ we find, when the transition from SG to LYF phase occurs. As was said before (formula (2)), it occurs at
 $\beta_1=\frac{\beta_c}{2}$, that corresponds to $d=19$. So besides $d=1$, there is another critical dimension for strings at $d=19$.
In [12] a version of string as a candidate for 3d Ising model has been proposed. It corresponds to $L_3$.\\
What is the cause of of the relation of REM to strings?
Both are representatives of complex systems. \\
Let us first set forth the basic principles underlying such systems in terms of REM, and then shortly review the complex systems.\\
It is striking at the phenomenological treatment of complex systems, that there are two kinds of variables,(spins and couplings in case of SG), as well as that at the definition the most important models prove to be locally symmetric.
Then we shall analyze the properties of REM in detail. 
In SG models one has spins s and couplings j. At intermediate stage of consideration of the theory of REM,
as well as of other models of SG, one comes across with the free energy at fixed couplings (the first form of free energy), and then after averaging the second form of free energy (the observable one) is encountered. Thus, in the beginning $Z(j)$ are calc
ulated and then the observed free energy for the system as a whole is calculated in the form of
average $<\ln Z(j)>$. In principle, instead of $\ln Z$ one can take the compressed form $Z(j)$ as $<Z(j)^{\mu}>,\mu<1$ [14]. In such a case $<Z(j)^{\mu}>$ corresponds to the partition of the system. In other SG-like models (glasses) one can single out the
 free energies not only at the level of the construction of theory, but also at the level of observables (macroscopic level). \\
REM is distinguished from the other
SG models by the fact, that the second free energy  has a special property, it
is equal to the logarithm of the number of couplings. The effective number in
the purest p-spin REM at $p=\frac{N}{2}$ is $2^N$, while the free
energy (have
been scaled to have transition at finite temperature) is proportional to N.\\
To proceed, let us introduce a concept of reflection as a mapping process (with compression in phase space) that conserves some essential properties of the initial phase space (the possibility to perform macroscopic work-the free energy in our case).\\
We define the thermodynamic reflection as the one in course of which the free energy is determined, and the strong reflection, as that where the free energy is the logarithm of the number of particles. \\
We define a REM like complex phenomena as the multiple reflection , the highest hierarchic level of which contains the strong reflection.\\
When at the definition of a model in the framework of certain theory we come upon strong reflection and local symmetry, we have sound reasons search for its relation with REM.
Let us define (in somewhat speculative manner) a concept of local symmetry (or of its loss) for the given single sample of SG (instead of an ensemble).\\
In case of pure REM nothing observable is changed in the PM
phase, when the signs of couplings are changed. As in case of SG phase it is impossible to identify such a symmetry,
it is assumed, that the system has lost the property of local symmetry.
We define the first phase of complex systems like this one as REM-SG phase.
Let us look for other phases of complexity. If in (3) our couplings in average have ferromagnetic part and there is
a representation for couplings 
\begin{equation}
\label{e19}
\begin{array}{l}
j_{i_1..i_p}= j^1_{i_1..i_p}+j^2_{i_1..i_p},\  \
j^1_{i_1..i_p}=J_0
\end{array}
\end{equation}
($j^2_{i_1..i_p} $ are normally distributed), in case of sufficiently high values of $J_0$ at low temperatures the system is
in the ferromagnetic phase with nonzero magnetization $<s_i>=m_i$. If our coupling $j^1$ are transformed at fixed values of $j^2$, then there takes place a symmetry
\begin{equation}
\label{e20}
m_i\to m_i\eta_i, \  \
j^1_{i_1..i_p}\to j^1_{i_1..i_p}\eta_{i_1}..\eta_{i_p}
\end{equation}
It is  important, that while the symmetry (20) is related to the initial one, it is nevertheless a different symmetry.
When the system moves from the boundary SG-FM inside the SG phase,
the  magnetization m and so the efficiency of symmetry (20)) decrease. We
denote this phase of complex system as REM-(SG-FM). In the following it will be
shown, that an analogous situation takes place in other complex systems too.
In the ordered (simple) phase the system has the property of local symmetry (possibly in equivalent formulation of
theory). This local symmetry may be either local gauge symmetry one or higher order symmetry (the order of symmetry
group is proportional to the volume of system or is higher by the order of magnitude). As the system moves to the 
higher complexity phase, it looses the property of local symmetry.\\
In principle, one can conceive other case (of weakly complex phase) resembling PM or LYF phases of REM. A bit is similar the transition SG-PM. This corresponds to $d=1$ case in strings.\\
Let us consider now the systems of different complexity from our (REM-oriented) viewpoint: presence of logically different degrees of freedom; local symmetry (under equivalent formulation of problem); number of reflections and availability of strong refle
ctions.\\
Turbulence. Is characterized by the presence of  Newtonian and
dissipative forces, local symmetry
(potentiality of velocity out of vortices) and  multiscaling. It
is a good candidate  for REM-(SG-FM).
 a short review of different complex systems.\\
2d conform theories ($c<1 $). Here  strong single reflection and local symmetry is observed. At the calculating of correlators REM-PM is encountered.\\
Liouville model. Describes the interaction of $c<1$ matter with 2d gravity. There are strong reflection and local symmetry. The model is dual with respect to REM. It may be in REM-PM and REM-SG phases.\\
Strings. There exist 2d gravity  and matter fields. After integration over  surfaces with different areas,
because of the zero mode integration of Laplacian one has double reflections including strong one.  REM-SG, REM-PM
and REM-LYF are possible here.\\
Abelian  SOC has some local (with very large group) symmetries, connected with different ways of toppling. 
Hence, besides the turbulence it is similar to  REM-(SG-FM) (Though the explicitly local symmetry is not observed at the 
formulation of theory).
Another probable representation of this class is edge of chaos,that is essential for the theory of evolution 
 according to S.Kauffman. Owing to the presence of multiscaling the availability of strong reflection is possible. Though the local symmetry here is not observed, REM-(SG-FM) is not excluded. It seems proper to remind one of highly interesting interpretat
ion of this phase (in case of REM) -the system is on the  threshold of errorless decoding.\\
Thus instead of probabilistic (entropic) approach to complexity I suggest (the free energy approach) to treat the complexity in terms of two criteria:\\
1. Degree of reflection\\
2. Extent  of symmetry breaking (loosing).\\
If the theory comprises free energy as a logarithmic of volume and a local symmetry is present at the definition of model,
 there is a valid chance of relation to REM.\\
In REM class of complex phenomena there is an information processing using the lowest level  language 
(any sequence of symbols is allowed, no correlations) due to local gauge invariance at the definition of model. In principle other classes of complex phenomena with higher complexity  languages may be allowed , as in case of protein folding.\\
I am grateful to  A'dam University and Fundacion Andes grant c-13413/1 for financial support.

\end{document}